\documentstyle[12pt,aasms4]{article}

\begin{document}

%\hskip 7cm {\bf Fermilab/Pub-00/050}

\title{The Effect of a Non-Thermal tail 
on the Sunyaev-Zel'dovich Effect in Clusters of Galaxies}

\author{P. Blasi$^*$, A. V. Olinto$^{\dag}$, and A. Stebbins$^*$}
\affil{$^*$ Fermi National Accelerator Laboratory, Batavia, IL 60510-0500}
\affil{$^{\dag}$Department of Astronomy \& Astrophysics, \& Enrico Fermi
Institute, University of Chicago, Chicago, IL 60637}

%\maketitle

\begin{abstract}
We study the spectral distortions of the cosmic microwave background radiation
induced by the Sunyaev-Zel'dovich (SZ) effect in clusters of galaxies when the
target electrons have a modified Maxwell-Boltzmann distribution with a
high-energy non-thermal tail.  Bremsstrahlung radiation from this type of
electron distribution may explain the supra-thermal X-ray emission observed in
some clusters such as the Coma cluster and A2199 and serve as an alternative to
the classical but problematic inverse Compton scattering interpretation. We
show that the SZ effect can be used as a powerful tool to probe the electron
distribution in clusters of galaxies and discriminate among these different
interpretations of the X-ray excess.  The existence of a non-thermal tail can
have important consequences for cluster based estimators of cosmological
parameters.
\end{abstract}

{\it Subject Headings:} Galaxies: clusters: general - Cosmology: theory
 
%\vskip 1cm

\section{Introduction}

Clusters of galaxies are powerful laboratories for
measuring cosmological parameters and for testing cosmological models of
the formation of  structure in the Universe.  These
associations of large numbers of galaxies are confined by a much greater
mass of dark matter, which also confines a somewhat smaller mass in very
hot gas. The galaxies and the gas are in rough virial equilibrium with the
dark matter potential well.  While initially clusters were investigated
through the observed dynamics of the galaxies they contain, in recent
decades much information has been gathered from studies of the gas,
primarily via X-ray observations of bremsstrahlung emission but also
through the Sunyaev-Zel'dovich (SZ) effect 
(Zeldovich \& Sunyaev 1969)\markcite{ZS69}. Interpreting these
observations requires a detailed understanding of the thermodynamic state
of the gas. With increasingly more sensitive measurements, the gas dynamics
should become clearer which would allow for a better understanding of the
structure and dynamics of clusters as well as their effectiveness as tests
of cosmological models.

Both the X-ray emission and the SZ effect are sensitive to the energy
distribution of the electrons.  It is usually assumed that in the intracluster
gas the electron energy distribution is described by a thermal
(transrelativistic Maxwell-Boltzmann) distribution function.  The typical
equilibration time for the bulk of this hot and rarefied electron gas is of
order $\sim 10^5$ years and is mainly determined by electron-electron Coulomb
scattering (electron-proton collisions are much less efficient). This time
rapidly increases when the electron energy becomes appreciably larger than the
thermal average, so that thermalization takes longer for higher energy
electrons. In the absence of processes other than Coulomb scatterings, the
electron distribution rapidly converges to a Maxwell-Boltzmann distribution.
However, the fact that the intracluster gas may be (and actually is often
observed to be) magnetized, can change this simple scenario: for instance,
cluster mergers can modify the electron distribution either by producing shocks
that diffusively accelerate part of the thermal gas, or by inducing the
propagation of MHD waves that stochastically accelerate part of the electrons
and heat most of the gas (Blasi 1999a)\markcite{Blasi99a}. 
Although the bulk of the electron
distribution is likely to maintain its thermal energy distribution, higher
energy electrons, more weakly coupled to the thermal bath, may acquire a
significantly non-thermal spectrum (Blasi 1999a)\markcite{Blasi99a}.

Until recently, X-ray observations could only probe energies below 
$\sim 10$ keV, where the observed radiation is consistent with
bremsstrahlung emission from the intracluster plasma with a thermal electron
distribution with temperatures in the $1-20$ keV range. The recent
detection of a hard X-ray component in excess of the thermal spectrum of the
Coma cluster (Fusco-Femiano et al. 1999)\markcite{Coma} may be the 
first indication that the particle
distribution  in (some) clusters of galaxies contains a significant
non-thermal component. Observations of Abell 2199 (Kaastra et al. 1999)
\markcite{A2199} show a
similar excess while no excess has been detected in  Abell 2319
(Molendi et al. 1999)\markcite{A2319}, thus, the source of this effect 
may not be universal. 

As argued above, the presence of magnetic fields in the intracluster gas allows
for acceleration processes that can modify the details of the heating
processes, so that the electron energy distribution may differ from a
Maxwell-Boltzmann. In this case, the bremsstrahlung emission from a modified
Maxwell-Boltzmann electron gas can account for the observed X-ray spectra, up
to the highest energies accessible to current X-ray observations
(Ensslin, Lieu \& Biermann 1999; Blasi 1999a)\markcite{ELB99,Blasi99a}. 
This model works as an alternative to the more
traditional interpretation based on the inverse Compton scattering (ICS)
emission from a population of shock accelerated ultra-relativistic electrons
(Volk \& Atoyan 1999)\markcite{VA99}. The ICS model has many 
difficulties such as the requirement that
the cosmic ray energy density be comparable to the thermal energy in the gas
(Ensslin et al. 1999; Blasi \& Colafrancesco 1999)
\markcite{ELB99,blasicola99}. This large cosmic ray energy
density might be hard to reconcile with  the nature of cosmic ray sources in
clusters (Berezinsky, Blasi \& Ptuskin 1997)\markcite{bbp97} and with 
gamma ray observations (Blasi 1999)\markcite{Blasi99}.
Moreover, the combination of  X-ray and radio observations within
the ICS model strongly indicates a very  low magnetic field,
$B\sim 0.1\mu G$, much lower than the values derived from Faraday rotation
measurements (Kim et al. 1990; Feretti et al. 1995)\markcite{kim,feretti},
which by themselves represent only lower limits to the field.

The best way to resolve the question of whether the observed hard X-rays are
due to ICS or are the first evidence for a modified thermal electron 
distribution in clusters is to probe directly such a distribution.  
We propose that this probe can be achieved by detailed observations of the
SZ effect, which is the change in brightness temperature of the cosmic
microwave background (CMB)  photons when they traverse a hot electron gas
such as the gas in clusters. We discuss the SZ effect in detail in the next
section, where the main results are also discussed. Additional implications
of the scenario proposed here are presented in section 3.

\section{The SZ effect as a probe of non-thermal processes}

In this section, we calculate the SZ effect
for a modified electron distribution, including a high energy tail.  We
follow the procedure outlined by Birkinshaw (1999)\markcite{SZreport}.

Photons of the CMB propagating in a gas of electrons are Compton scattered and
their energy spectrum is modified. As long as the 
center-of-mass energy of the collision is less than $m_{\rm e}c^2$, the
scattering is accurately described by the Thomson differential cross-section.
For CMB photons at low redshift this only requires that the electron
energy in the cosmic rest-frame be less than $\sim1$\,TeV.  For scattering
of a photon with initial frequency $\nu_{\rm i}$, off an isotropic
distribution of electrons each with speed $v$, the probability distribution
of the scattered photon having frequency $\nu_{\rm i}(1+\Delta)$ is
(Stebbins 1997)\markcite{CMBRspectrum}
\begin{equation}
P(\Delta,\beta)\,d\Delta=
{\overline{F}(\Delta,\beta\,{\rm sgn}(\Delta))\over(1+\Delta)^3}\,d\Delta \ ,
\qquad
\Delta\in\left[-{2\beta\over1+\beta},{2\beta\over1-\beta}\right]
\label{eq:Pdb}
\end{equation}
where $\beta={v\over c}$ and 
\begin{eqnarray}
&&\overline{F}(\Delta,b)=
\Biggl|
 {3(1-b^2)^2(3-b^2)(2+\Delta)\over16b^6}\,\ln{(1-b)(1+\Delta)\over1+b}
 +{3(1-b^2)(2b-(1-b)\Delta)\over32b^6(1+\Delta)}                   \cr
&&\hskip75pt
   \times\left(4(3-3b^2+b^4)+2(6+b-6b^2-b^3+2b^4)\Delta
                                  +(1-b^2)(1+b)\Delta^2\right)\Biggr|\ .
\end{eqnarray}
If instead of a fixed speed, we consider the scattering off electrons
with a distribution of speeds,
${\rm p}(\beta)\,d\beta$, the distribution of $\Delta$ after one scattering
is 
%\begin{equation}
$P_1(\Delta)=\int_{|\Delta|/(2+\Delta)}^{1} d\beta\,{\rm
p}(\beta)\,P(\Delta,\beta)\ .$
%\label{eq:P1}
%\end{equation}
This expression can be easily applied to determine the change in the
spectrum of the CMB  as seen through the hot gas in a cluster of
galaxies.  Since clusters have a small optical depth to Compton scattering
($\sim10^{-2}$), the fraction of photons which are scattered is given by the
optical depth,
$\tau_{\rm e}=\sigma_{\rm T} N_{\rm e}$, where $N_{\rm e}$ is the projected
surface density of free electrons. 
The change in brightness of the CMB at frequency $\nu$ 
due to the SZ effect is then given by
\begin{equation}
\Delta I(\nu) = \frac{2h\,\nu^3}{c^2} \tau_e\int_{-1}^{+\infty}d\Delta\,
P_1(\Delta)\,\left[{(1+\Delta)^3\over e^{(1+\Delta)\,x}-1} -
                   { 1          \over e^x              -1}\right],
\label{eq:DI}
\end{equation}
where $x\equiv{h\nu\over k_{\rm B}T_{\rm CMB}}$, $T_{\rm CMB}$ is the CMB
temperature at the present epoch, and $k_{\rm B}$ is   
Boltzmann's constant.  It
is conventional in CMB studies to use the change in the thermodynamic 
brightness temperature rather than the change in brightness, the former
being given by 
\begin{equation}
\frac{\Delta T}{T_{\rm CMB}}={(e^x-1)^2\over x^4\,e^x}\,{\Delta I\over I_0}
\label{eq:DeltaT}
\end{equation}
where $I_0\equiv{2(k_B T_{\rm CMB})^3\over(hc)^2}$.

For very non-relativistic electrons, $P_1(\Delta)$ is narrowly peaked and can
be accurately estimated via a 1st order Fokker-Planck approximation. This gives
the classical formula (Zeldovich \& Sunyaev 1969)\markcite{ZS69}
%\begin{equation}
$\frac{\Delta T}{T_{\rm CMB}}=y\,\left(x{e^x+1\over e^x-1}-4\right),$
%\label{eq:classicalSZ}
%\end{equation}
where $y={1\over3}\tau_{\rm e}\,\langle\beta^2\rangle$.  In this limit the
shape of the spectral distortion yields no useful information, only the
amplitude, $y$, is interesting but it depends only on the 2nd moment of  ${\rm
p}(\beta)$. Fortunately the gas in rich clusters is hot enough for relativistic
corrections to become important, leading to deviations from this  classical
formula at the  $\sim$10\% level 
(Birkinshaw 1999, Rephaeli 1995, Stebbins 1997, Challinor \& Lasenby 1998,
Itoh, Kohyama \& Nozawa 1998)
\markcite{SZreport,CMBRspectrum,Rephaeli95,CL98,IKN98}.  
Through these relativistic
corrections, changes in the electron energy distribution can be measured by the
modified shape of the SZ spectrum, hence the shape of the SZ effect can be used
to differentiate between thermal and non-thermal models. Even without spectral
information, non-thermality can be inferred by the comparison of the X-ray
flux and temperature with the amplitude of $\Delta T_{\rm SZ}$, however
this requires a detailed model of the density structure of the cluster
since the SZ effect and bremsstrahlung emission scale differently with
density.

The SZ effect is usually computed assuming a thermal ${\rm p}(\beta)$, but
here we include the effect of a non-thermal tail. We adopt the model for the
distribution function used by Ennslin et al. (1998)\markcite{ELB99} 
which fits both the non-thermal
hard X-ray data and the thermal soft X-ray data.  In particular, a thermal
distribution for momenta smaller than $p^*$ 
($\equiv m_{\rm e}c\,\beta^*\gamma^*$)
is matched to a power law distribution in momentum above $p^*$, and cutoff at
momentum $p_{\rm max}$ ($\equiv m_{\rm e}c\,\beta_{\rm max}\gamma_{\rm max}$)
i.e.
\begin{equation}
{\rm p}(\beta)={C\gamma^5\beta^2\over
\Theta\,K_2({1\over\Theta})}\times\left\{
\matrix{{\rm exp}(-{\gamma \over \Theta})&       & \quad\beta\in[0,\beta^*] \cr
        {\rm exp}(-{\gamma^*\over\Theta})&
 \hskip-10pt({\beta^*\gamma^*\over\beta\gamma})^{\alpha+2}
                                     &\quad\beta\in[\beta^*,\beta_{\rm max}]\cr
                0            &       &\quad\beta\in[\beta_{\rm max},1)
        }\right.\ .
\end{equation}
Here $\gamma={1\over\sqrt{1-\beta^2}}$,
$\gamma^*={1\over\sqrt{1-{\beta^*}^2}}$, $\Theta={kT\over m_e c^2}$ gives the
temperature of the low energy thermal distribution, and $C$ ($\approx1$)
normalizes the function to unit total probability. For instance, in the model
proposed by Blasi (1999a)\markcite{Blasi99a}, a cutoff at 
$\beta_{\rm max}\gamma_{\rm max}\sim
1000$ arises naturally and insures that the electrons in the tail do not affect
the synchrotron radio emission.
For $\gamma_*\gg1$ one finds $C=0.982$, indicating that only 1.8\% of the
electrons are in the non-thermal tail, however the electron kinetic energy is
increased by 73\% and the electron pressure by 48\%, so the hydrodynamical
properties of the gas can be greatly influenced by the non-thermal component.

The bremsstrahlung emissivity is given by
%\begin{equation}
$
q_{brem}(k_\gamma) = n_{gas} \int dp ~n_e(p) ~v(p) ~\sigma_B(p,k_\gamma),
\label{eq:brem}
$
%\end{equation}
where $n_{gas}$ is the gas density in the cluster, $v(p)$ is the velocity
of an electron with momentum $p$ and $k_\gamma$ is the photon momentum. The
bremsstrahlung cross section, $\sigma_B$, is taken from 
(Haug 1997)\markcite{haug97}. 
%In deriving eq. (\ref{eq:brem}), 
We assumed for simplicity that the cluster has
constant density and temperature, but our results  can be easily
generalized to the more realistic spatially varying case.

As shown by Ensslin et al. (1999)\markcite{ELB99}, there is a 
wide region in the $p^*$-$\alpha$
parameter space that matches the observations. We choose the values
$\beta^*\gamma^*=0.5$ and $\alpha=2.5$ that provide a good fit to the overall
X-ray data, as shown in Fig.~1, where the thermal component has a temperature
$T=8.21$ KeV. The data points are from BeppoSAX 
(Fusco-Femiano 1999)\markcite{personal} observations,
while the thick curve is the result of our calculations for a suitable choice
of the emission volume.

The basic question that we want to answer is whether the non-thermal tail
in the electron distribution produces distortions in the CMB radiation that
can be distinguished from the thermal SZ effect. To answer this question,
we calculate the SZ spectrum using eq.~(\ref{eq:DI}),
plotting the results in Fig.~2, for a thermal model and two non-thermal
models, each based on Coma.  There is an appreciable difference between the
curves, as large as $\sim 60\%$ at high frequencies ($x>5$). At low
frequencies ($x<1.7$), the region currently probed by most SZ observations, the
relative difference is at the level of $\sim10-20\%$. 

To establish the existence of a non-thermal contribution to the SZ effect, say
in Coma, one should measure $\Delta T$ at three or more frequencies. While
$T_{\rm e}$ is well constrained by X-ray measurements, $\tau_{\rm e}$ is not,
and in addition the SZ distortion is contaminated by a frequency independent
constant, $\Delta T_{\rm CMBR}+\Delta T_{\rm kSZ}$, i.e. the sum of the
background primordial CMBR anisotropy, and the kinematic SZ effect caused by a
line-of-sight velocity, $v_{\rm c}$, in the CMBR frame.  Two measurements are
required to determine these unknowns before one is able to detect
non-thermality.  In fig.~3 we estimate the difference in $\Delta T$ for a
thermal and non-thermal spectrum after allowing for these unknowns.  The
residual spectral difference remain both at low and high frequencies, and might
be accessible by ground observation.  From space a non-thermal signature should
be detectable by the Planck Surveyor, but not by MAP, mainly due to sensitivity
and beam dilution rather than frequency coverage.

Of particular interest observationally is the frequency of the zero SZ spectral
distortion, $x_0$, defined by $\Delta I(x_0)=\Delta T(x_0)=0$.  Measuring the
difference in the CMB flux on and off the cluster near the zero allows
the measurement of small deviations from the classical behaviour with
only moderate requirements on the calibration of the detector, and is very
sensitive to $v_{\rm c}$. For a thermal plasma 
(Birkinshaw 1999)\markcite{SZreport},
%\begin{equation}
$x_0=3.830\,\left(1+1.13\,\Theta+1.14\,{\beta_{\rm c}\over\Theta}\right)
                                        +{\cal O}(\Theta^2,\beta_{\rm c}^2)$,
%\label{eq:x0}
%\end{equation}
where $\Theta={kT_e\over m_{\rm e} c^2}$, $\beta_{\rm c}={v_{\rm c}\over c}$.
This equation is no longer valid for a non-thermal electron distribution.
For our canonical parameters, no cutoff, and $v_{\rm c}=0$, we find that $x_0$
is shifted to 3.988, the same as would be obtained for a thermal distribution
with an unreasonably large temperature of 18.62\,keV, and $v_{\rm c}=0$, or
with the ``correct'' temperature (8.21\,keV) and  $v_{\rm c}=111\,$km/sec.
Even with our non-thermal tail, it is the velocity which mostly determines the
value of $x_0$, although the non-thermal electrons can bias the $v_{\rm c}$
determinations by $\sim+100$\,km/sec (i.e. away from the observer).

\section{Other Implications}

In this section we mention other important consequences of the existence of a
non-thermal electron distribution.  As noted above, the non-thermal component
might correspond to only a few percent in additional electrons which  do
not contribute significantly to the nearly thermal 1-10\,keV X-ray
emission, while at the same time the electron pressure may be increased
by nearly a factor of two (we have no evidence whether there is similar
increase in the ion pressure).  Many cluster mass estimates which are
based on X-ray observations, use the hydrostatic relation $M_{\rm
c}\propto\nabla p/\rho$, and if the pressure has been significantly
under-estimated due to non-thermal electrons the cluster mass would also
be underestimated. Cluster masses play an important role in normalizing
the amplitude of inhomogeneities in cosmological models, and the
non-thermal electron populations may lead to an underestimate in this
cluster normalization.  The baryon fraction in clusters have also been
used as an indicator of the universal baryon-to-mass ratio, $\Omega_{\rm
b}/\Omega_{\rm m}$. If a cluster mass is underestimated due to
non-thermal electrons then the cluster baryon fraction will be
overestimated.  Note that the Coma cluster, which does have a non-thermal
X-ray excess, has played a particularly important role in cluster
$\Omega_{\rm b}/\Omega_{\rm m}$ estimates 
(White et al. 1993)\markcite{WNEF93} although optical mass
estimates are also used here. 
These cosmological consequences would be true even if the 
excess pressure was provided by a population of relativistic cosmic rays,
as discussed by Ensslin et al. (1997).

Other implications are instead peculiar to the non-thermal tail scenario:
using a combination of X-ray and SZ measurements, clusters have been used to 
estimate Hubble's constant, $H_0\propto I_{\rm X}/(\Delta T_{\rm SZ})^2$
(Birkinshaw 1999)\markcite{SZreport}.  
We have shown that a non-thermal electron distribution
generally increases $\Delta T_{\rm SZ}$ for fixed $\tau_{\rm e}$ and $\Theta$,
and therefore one should use a larger proportionality constant when
non-thermal electrons are present.  Therefore cluster estimates of $H_0$
without taking into account a non-thermal electron distribution would
under-estimate $H_0$.

If our model of the non-thermal tail held universally then naive estimates of
$M_{\rm c}$, $\Omega_{\rm b}/\Omega_{\rm m}$, and $H_0$ should be respectively
adjusted upward, downward, and upward by 10's of percent.  However estimates of
cosmological parameters using clusters generally make use of measurements of an
ensemble of clusters.  Supra-thermal X-ray emission does appear in two of three
clusters, but the statistics are not good enough for an accurate prediction of
how frequently a non-thermal electron distribution might be present in a sample
of clusters.  Therefore the overall bias introduced in parameter estimates
is necessarily uncertain.  In any individual cluster the bias in a parameter
estimator will depend on the spatial distribution of the non-thermal electrons,
which is also uncertain and not well-constrained by present hard X-ray
measurements.  The important point is that the magnitude of cosmological
parameter mis-estimation might be quite large.

Confirmation or refutation of the hypothesis that the X-ray excess is due to a
non-thermal tail will have important consequences not only for the
understanding of cluster structure but for cosmology as well.  We argue that SZ
measurements are the best way to test this hypothesis, and that this is within
the capabilities of present technology.

\section{Acknowledgements}
We are grateful to M. Bernardi for a useful discussion. The work of P.B. and
A.S. was funded by the DOE and the NASA grant NAG 5-7092 at Fermilab. The work
of A.V.O. was supported in part by the DOE through grant DE-FG0291 ER40606, and
by the NSF through grant AST-94-20759.

\newpage

\figcaption[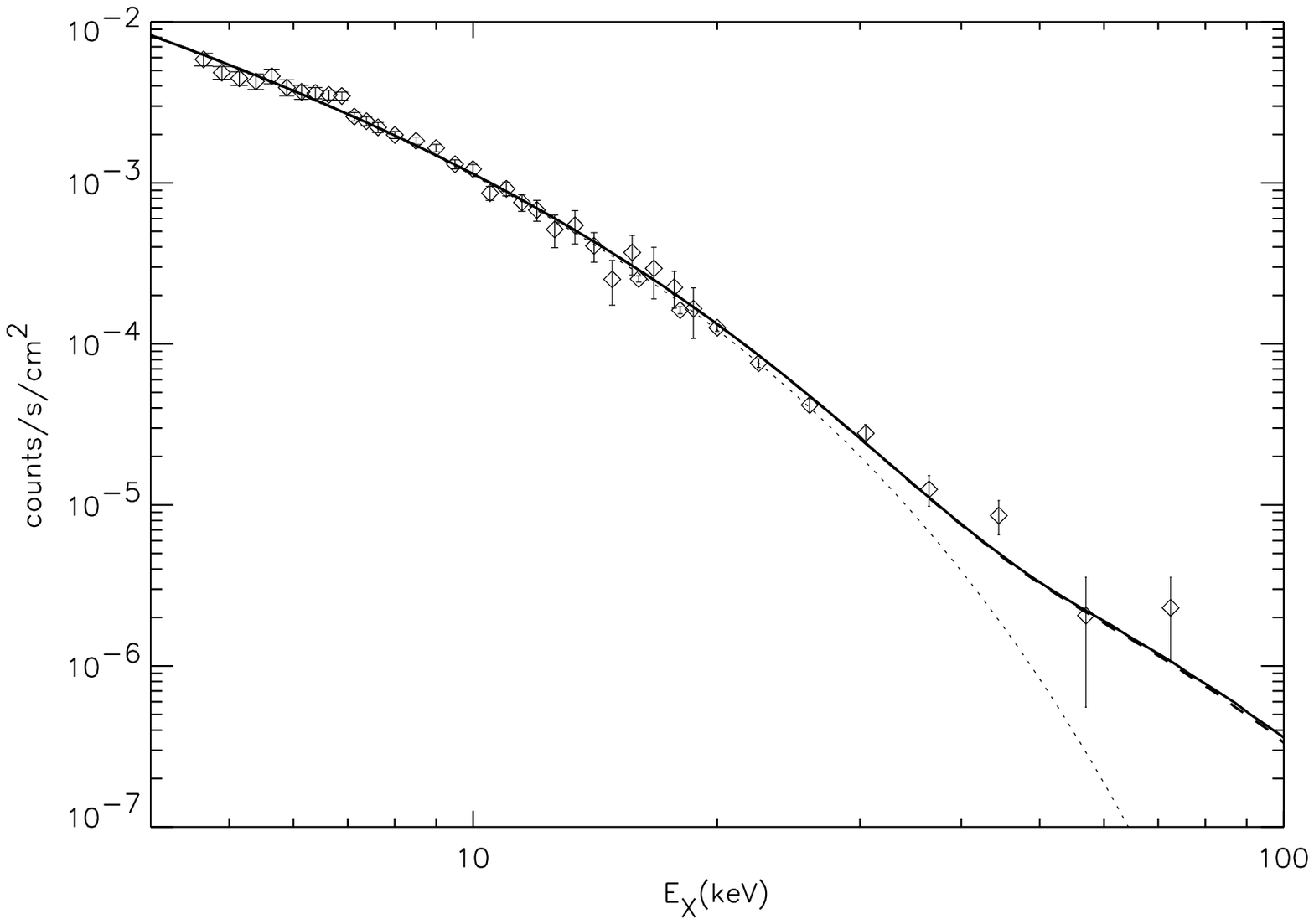]{
X-ray emission from the Coma cluster for the modified 
electron distribution adopted in our calculations. The solid
line is for $\beta_{max}\gamma_{max}=\infty$ while the dashed line 
is for $\beta_{max}\gamma_{max}=1$.
The dotted line is the contribution expected from a purely thermal 
electron distribution.}

\figcaption[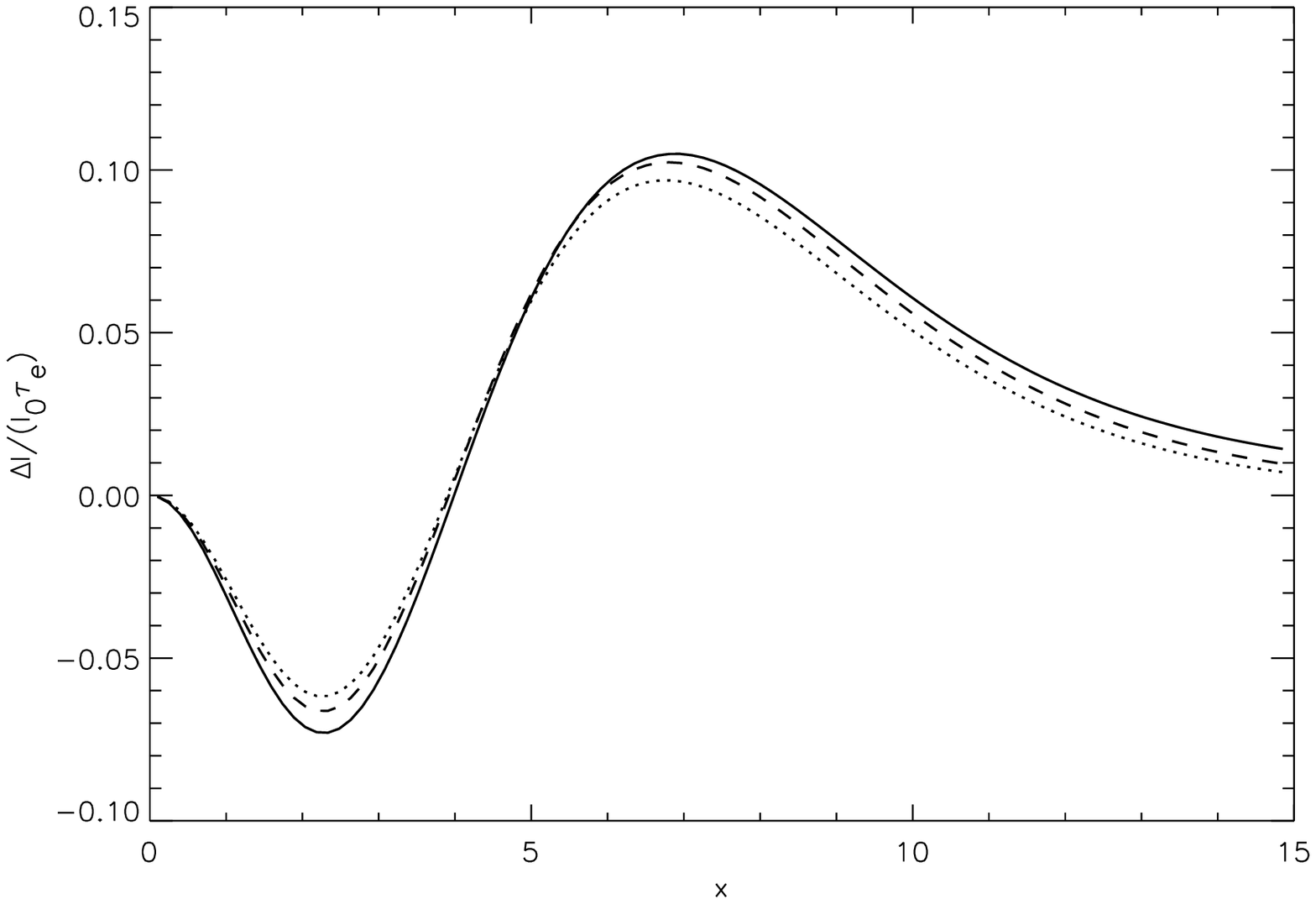]{
Fractional brightness change normalized to the
opacity as a function of the dimensionless frequency $x$. 
Lines are labeled as in Fig. 1.}

%\figcaption[DT.ps]{
%Fractional temperature change as defined in 
%eq. (\ref{eq:DeltaT}). The lines are labeled as in Fig. 1.}

\figcaption[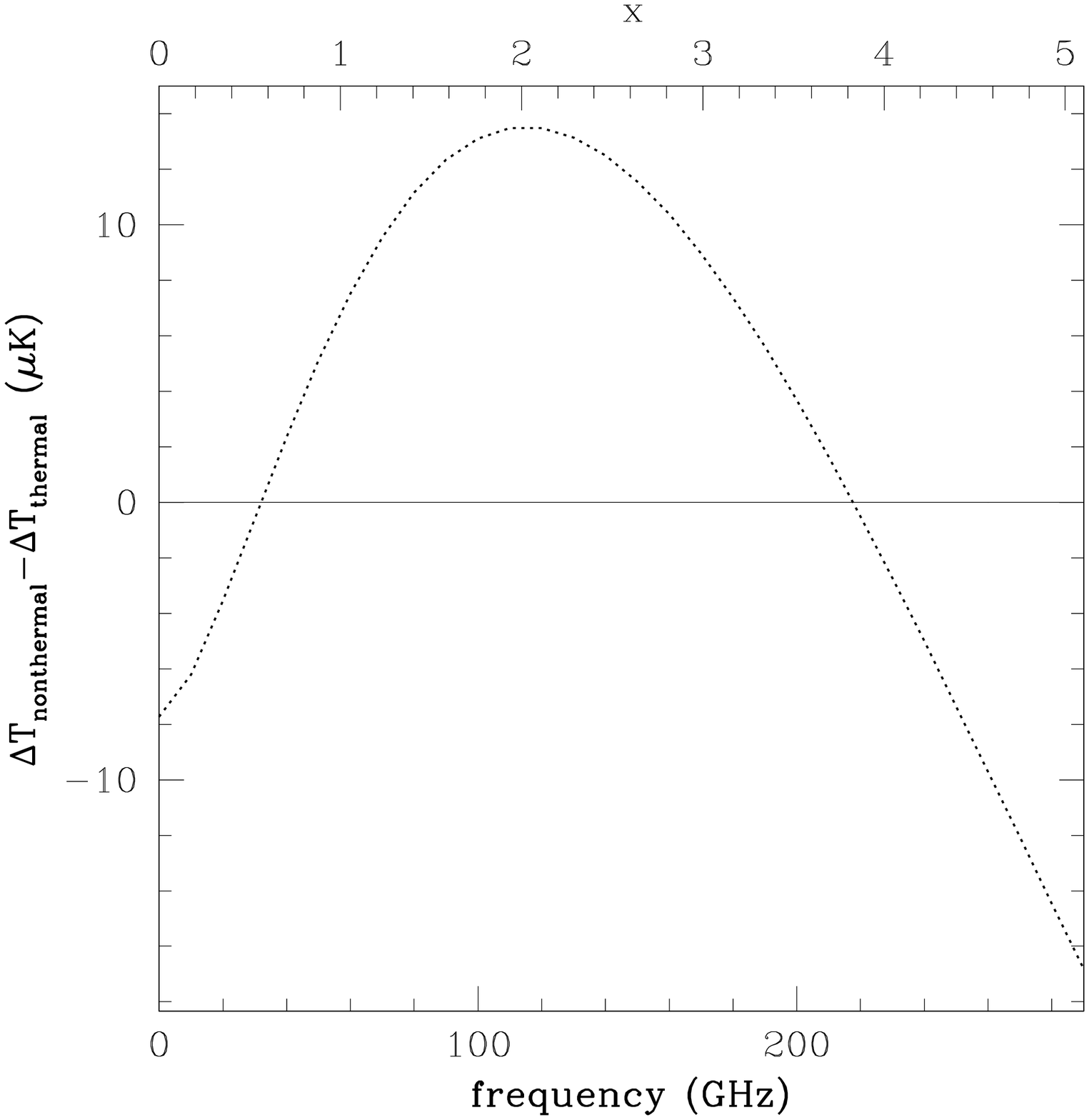]
{Here is plotted the difference in the Coma cluster's central
$\Delta T$ between our non-thermal model 
($\beta_{max}\gamma_{max}=\infty$) and
a thermal model. Both have $T_{\rm e}=8.21$ keV to fit the X-ray data, 
but the
non-thermal model has, relative to the thermal model, $\tau_{\rm e}$ 
decreased
by 13\% and $\Delta T_{\rm CMBR}+\Delta T_{\rm kSZ}$ decreased by 
$17\mu$K.
Parameters were chosen so that both match the inferred central 
decrement of
$-550\mu$K at 32 GHz (Herbig et al. 1995) and also 
match at 218 GHz, near $x_0$.
This exemplifies how one might use sensitive three channel 
data (at
32 GHz, 218 GHz, and a third frequency, say 100 GHz) to detect 
a non-thermal
component in the electron distribution.}

\end{document}